\documentclass[10pt,twocolumn,conference]{IEEEtran}
\usepackage{setspace}
\usepackage{clrscode}
\usepackage{pict2e}

\usepackage{amsmath}
\usepackage{amssymb}
\usepackage[dvips]{graphicx}
\usepackage{epsfig}
\usepackage{algorithm}
\usepackage{algorithmic}
\usepackage{caption}
\usepackage{amsthm}
\usepackage{subcaption}

\usepackage[ps2pdf,
bookmarks=false,
bookmarksnumbered=false, 
bookmarksopen=false, 
colorlinks=false]{}

\newcommand{\figsize}{0.39}

\makeatletter

\newcommand{\Rmnum}[1]{\expandafter\@slowromancap\romannumeral #1@}
\makeatother

\newcommand{\agg}{\text{agg}}
\newcommand{\G}{\text{G}}
\newcommand{\g}{\text{g}}
\newcommand{\avg}{\text{avg}}

\newcommand{\pG}{p_{\text{\G}}}
\newcommand{\BE}{\text{BE}}
\newcommand{\SNR}{\text{SNR}}

\begin{document}

%
\title{Average Error Probability Analysis in mmWave Cellular Networks}

\author{\authorblockN{Esma Turgut and M. Cenk Gursoy}
\authorblockA{Department of Electrical Engineering and Computer Science\\
Syracuse University, Syracuse, NY 13244\\ Email:
eturgut@syr.edu, mcgursoy@syr.edu}}

\maketitle

\begin{abstract}
In this paper, a mathematical framework for the analysis of average symbol error probability (ASEP) in millimeter wave (mmWave) cellular networks with Poisson Point Process (PPP) distributed base stations (BSs) is developed using tools from stochastic geometry. The distinguishing features of mmWave communications such as directional beamforming and having different path loss laws for line-of-sight (LOS) and non-line-of-sight (NLOS) links are incorporated in the average error probability analysis. First, average pairwise error probability (APEP) expression is obtained by averaging pairwise error probability (PEP) over fading and random shortest distance from mobile user (MU) to its serving BS. Subsequently, average symbol error probability is approximated from APEP using the nearest neighbor (NN) approximation. ASEP is analyzed for different antenna gains and base station densities. Finally, the effect of beamforming alignment errors on ASEP is investigated to get insight on more realistic cases.
\end{abstract}
%

\thispagestyle{empty}


\section{Introduction}
Demand for mobile data and use of smart phones have been increasing very rapidly in recent years. According to the UMTS traffic forecasts, 1000 fold increase in mobile data traffic is predicted by the year 2020 \cite{UMTS}. In another estimate, more than 50 billion devices may be connected wirelessly by 2020 which may cause a capacity crisis \cite{Ericsson}. To meet this increasing demand, moving to new frequency bands becomes a necessity. Therefore, the large available bandwidth at millimeter wave (mmWave) frequency bands, between 30 and 300 GHz, becomes a good candidate for the fifth generation (5G) cellular networks and has attracted considerable attention recently \cite{Rappaport}, \cite{Andrews}, \cite{Rangan}, \cite{Ghosh}.


Evaluating the system performance of mmWave cellular networks is a crucial task in order to understand the network behavior. There are several recent studies which analyze the coverage probability and average rate in mmWave cellular networks using results from stochastic geometry and the theory of point processes for different base station(BS)-user associations \cite{Bai2}, \cite{Marco2}, \cite{Marco4}. Stochastic geometry is a commonly used powerful mathematical tool to evaluate the average network performance of spatially distributed nodes \cite{Baccelli}. Poisson point process (PPP) is a widely used model in wireless networks in general and in cellular networks in particular due to its analytical tractability. However, average error probability in PPP-based cellular networks has only been barely analyzed in the literature. For instance, there is work focusing on the computation of average symbol error probability (ASEP) in the presence of Poisson field interferers (see e.g., \cite{Win}). However, none of them are applicable to cellular networks since the BS to mobile user (MU) cell association is generally not considered. In \cite{marco}, a mathematical framework to compute the ASEP in cellular networks, where the BS locations are modeled as independent homogeneous PPPs, is established for the first time. Their approach is based on the shortest BS-to-MU distance cell association criterion, which guarantees that the interfering BSs are located farther than the serving BS, so it is applicable to cellular networks. However, to the best of our knowledge, average error probability analysis has not been conducted for mmWave cellular networks yet.

In this paper, we follow a similar approach as in \cite{marco} to develop a mathematical framework for the computation of ASEP in downlink mmWave cellular networks. First, average pairwise error probability (APEP) is calculated by averaging PEP over fading and random shortest distance from MU to serving BS. Then, ASEP is found using the nearest neighbor (NN) approximation. The main contribution of this paper is the combination of the characteristic features of mmWave communications with the proposed mathematical framework in \cite{marco}. One distinguishing feature of mmWave cellular communication is the directional beamforming at the transmitter and receiver, which provides an array gain to mitigate the effect of path loss. In this paper, sectored directional antenna model is used to find the effective antenna gain similar to \cite{Bai2}, \cite{Marco2}, \cite{Bai3}. First, perfect beam alignment is assumed between the MU and the serving BS. Then, the effect of beamsteering errors is investigated. Another distinct feature of mmWave communication is that a BS can be in the line-of-sight (LOS) or in the non-line-of-sight (NLOS) of MU and different path loss laws are applied for LOS and NLOS links. Here, we adopt the equivalent LOS ball model in \cite{Bai2} to determine whether a BS is LOS or NLOS.

\section{System Model} \label{sec:system_model}
In this section, we introduce our system model for the downlink mmWave cellular  network consisting of BSs distributed according to some homogeneous PPP $\Psi$ of density $\lambda$ in the Euclidean plane. Without loss of generality, we consider that a typical MU is located at the origin. A shortest distance cell criterion is assumed, i.e., MU is served by the nearest BS which is denoted by $\text{BS}_0$.
The distance from the $i$th BS to the MU is denoted by $r_i$ for $i \in \Psi$. Thus, the distance between the MU and serving BS ($\text{BS}_0$) is $r_0$ which is a random variable (RV) with PDF $f_{r_0}(\xi)=2\pi \lambda \xi \exp\{-\pi\lambda\xi^2\}$ \cite{Andrews2}. The set of interfering BSs $i \in \Psi -{\text{BS}_0}$ is still a homogeneous PPP, denoted by $\Psi^{(\setminus0)}$, according to the Slivnyak-Mecke's Theorem \cite{Baccelli}. We assume that all the interfering BSs are transmitting in the same frequency band as the serving BS (full frequency reuse), therefore $\Psi^{(\setminus0)}$ has density $\lambda$ as well.

We have the following two assumptions in the construction of the system model.

\textbf{Assumption 1:} Antenna arrays at both the BSs and MU are used to perform directional beamforming such that the main lobe is directed towards the dominant propagation path while smaller sidelobes direct energy in other directions. For tractability in the analysis, antenna arrays are approximated by a sectored antenna model, in which the array gains are assumed to be equal to a constant $M$ for all angles in the main lobe and another smaller constant $m$ in the side lobe \cite{Hunter}. The MU and serving BS, $\text{BS}_0$, are assumed to have perfect beam alignment and therefore have an antenna gain of $MM$. Also, the beam direction of the MU and each interfering BS can be modeled as a uniform random variable on $[0,2\pi]$. Therefore, the effective antenna gain is a discrete RV described by
\begin{equation}
    \G_i=\left\{
                \begin{array}{ll}
                  MM & \text{with prob.} \; p_{MM}=(\frac{\theta}{2\pi})^2\\
                  Mm & \text{with prob.} \; p_{Mm}=2\frac{\theta}{2\pi}\frac{2\pi-\theta}{2\pi}\\
                  mm & \text{with prob.} \; p_{mm}=(\frac{2\pi-\theta}{2\pi})^2
                \end{array}
              \right., \label{eq:antennagains}
\end{equation}
where $\theta$ is the beam width of the main lobe.

\textbf{Assumption 2:} A BS can be either LOS or NLOS BS to the MU according to the $LOS$ $probability$ $function$ $p(r)$ which is the probability that a link of length $r$ is LOS. Using field measurements and stochastic blockage models, $p(r)$ can be formulated as $e^{-\beta d}$ where decay rate $\beta$ depends on the building parameter and density \cite{Bai1}. $LOS$ $probability$ $function$ $p(r)$ can be approximated by a step function in order to simplify the analysis. In this approach, the irregular geometry of the LOS region is replaced with its equivalent LOS ball model with radius $R_B$ \cite{Bai2}. A BS is a LOS BS to the MU if it is inside the ball, otherwise it is a NLOS BS. Different path loss laws are applied to LOS and NLOS links. Thus, the path-loss exponent on each interfering link can be expressed as follows:
\begin{equation}
    \alpha_i=\left\{
                \begin{array}{ll}
                  \alpha_L \quad \text{if} \; r \le R_B\\
                  \alpha_N \quad \text{if} \; r > R_B
                \end{array}
              \right.,  \label{eq:Pathloss}
\end{equation}
where $\alpha_L$ and $\alpha_N$ are the LOS and NLOS path-loss exponents, respectively.

By combining these two assumptions with the described network model above, the received signal at the MU can be written as,
\begin{equation}
y=\underbrace{\sqrt{G_0 E_0}r_0^{-\alpha_{L}}h_0s_0}_{x}+\underbrace{\sum_{i \in \Psi^{(\setminus0)}} \sqrt{G_i E_0}r_i^{-\alpha_{i}}h_is_i}_{I_{\agg}}+n.  \label{eq:receivedsig}
\end{equation}
where $x$ is the signal arriving from the serving BS to MU, $I_{\agg}$ is the aggregate network interference, and $n$ is the Gaussian distributed noise component with zero mean and variance $N_0$. Moreover, $G_0$ is the effective antenna gain of the $\text{BS}_0$-MU link and it is assumed to be equal to $MM$, $E_0$ is the BSs' transmit-energy per transmission, $\alpha_{L}$ is the LOS path-loss exponent of the $\text{BS}_0$-MU link, $s_0=a_0\exp{\{j\theta_0\}}$ is the information symbol transmitted by $\text{BS}_0$ with amplitude $a_0$ and phase $\theta_0$, $h_0=|h_0|\exp{\{j\phi_0\}}$ is the fading coefficient in the $\text{BS}_0$-MU link where $|h_0|^2$ is an exponential RV with parameter $\sigma_0=\mathbb{E}{|h_0|^2}=1$ and the phase $\phi_0$ is a uniformly distributed RV in the range $[0,2\pi)$. A similar notation is used for $I_{\agg}$, but note that the effective antenna gain $G_i$ and path loss exponent $\alpha_i$ are different for different interfering links as described in (\ref{eq:antennagains}) and (\ref{eq:Pathloss}), respectively. Also, considering phase modulation, we assume that $a_0=a_i=1$ for $i \in \Psi^{(\setminus0)}$.

At the MU, an interference-unaware maximum-likelihood (ML) demodulator is used as in \cite{marco}, which can be formulated  as \cite{Simon}
 \begin{equation}
 \hat{s}_0= \arg\min_{\tilde{s}_0}\{D(\tilde{s}_0)=|y-\sqrt{G_0 E_0}r_0^{-\alpha_{L}}h_0\tilde{s}_0|^2\}.  \label{eq:ML}
 \end{equation}
Inserting (\ref{eq:receivedsig}) into (\ref{eq:ML}) and neglecting some irrelevant constants after algebraic manipulations, we can express the decision metric as
\begin{align}
D(\tilde{s}_0)&\propto r_0^{-2\alpha_{L}}G_0 E_0|\Delta_{s,\tilde{s}}|^2|h_0|^2 \nonumber \\
&+2r_0^{-\alpha_{L}}\sqrt{G_0 E_0}\text{Re}\{(I_{\agg}+n)|h_0|\exp{\{-j\phi_0\}}\Delta_{s,\tilde{s}}^\ast\}, \label{eq:DecMet}
\end{align}
where $\Delta_{s,\tilde{s}}=s_0-\tilde{s}_0$, $I_{\agg}=\sum_{i \in \Psi^{(\setminus0)}} \sqrt{G_i E_0}r_i^{-\alpha_{i}}h_is_i$ and $n\sim C \mathcal{N}(0,N_0)$. Since the effective antenna gain between the MU and each interfering BS is modeled as an independent RV, we can employ the thinning property of PPP to split the aggregate network interference $I_{\agg}$ into 6 independent PPPs as follows \cite{Bai3}:
\begin{align}
I_{\agg}&=(I_{\Psi_{\text{LOS}}}^{MM}+I_{\Psi_{\text{NLOS}}}^{MM})+(I_{\Psi_{\text{LOS}}}^{Mm}+I_{\Psi_{\text{NLOS}}}^{Mm})+(I_{\Psi_{\text{LOS}}}^{mm}+I_{\Psi_{\text{NLOS}}}^{mm})  \nonumber \\
&=\sum_{G \in \{MM,Mm,mm\}}(I_{\Psi_{\text{LOS}}}^{G}+I_{\Psi_{\text{NLOS}}}^{G}),  \label{eq:6PPP}
\end{align}
where each interfering BS is either a LOS or NLOS BS and the superscripts represent the discrete random antenna gain defined in (\ref{eq:antennagains}). According to the thinning theorem, each independent PPP has a density of $\lambda p_G$ where $p_G$ is given in (\ref{eq:antennagains}) for each antenna gain $G \in \{MM,Mm,mm\}$.

\section{Average Error Probability Analysis} \label{sec:error_prob_analysis}
In this section, we investigate the error performance of a downlink mmWave cellular network. The first step in obtaining an approximation of the average error probability is to compute the pairwise error probability (PEP) associated with the transmitted symbols. Hence, initially we derive an expression for PEP, conditioned on fading gain ($|h_0|$) and random shortest distance of the MU-serving BS link ($r_0$), in terms of the characteristic function (CF) of the aggregate network interference and the noise. A closed-form expression is determined for the CF of the aggregate network interference for PPP distributed BSs. Then, APEP is computed by averaging the conditional PEP over fading and the position of the serving BS. Finally, ASEP is approximated from APEP using the NN approximation.

\subsection{Derivation of Pairwise Error Probability}
The PEP is defined as the probability of deciding in favor of $\hat{s}_0$ when actually $s_0$ is transmitted. It is assumed that these two symbols are the only two symbols in the signal-constellation, and therefore decision is made strictly between these two symbols. Using the decision metric in (\ref{eq:DecMet}), PEP conditioned on $|h_0|$ and $r_0$ can be computed as
\begin{equation}
P\{s_0 \to \hat{s}_0||h_0|,r_0 \}=P \{ D(\tilde{s}_0 = \hat{s}_0) < D(\tilde{s}_0 =s_0) \}.  \label{eq:PEPdef}
\end{equation}
When $\tilde{s}_0 =s_0$, $\Delta_{s,\tilde{s}}=s_0-\tilde{s}_0$ becomes zero by definition. As a result, $D(\tilde{s}_0 =s_0)$ is zero. Let $U=I_{\agg}+n$. Note that $U$ is a circularly symmetric RV. Thus, $U\exp{\{j\phi_0\}}\text{arg}\{\Delta_{s,\hat{s}}^\ast\} \stackrel{d}{=}U$ \cite{Samoradnitsky}. Thus, PEP can be computed as follows:
\begin{align}
&P\{s_0 \to \hat{s}_0||h_0|,r_0 \}=P \{ D(\tilde{s}_0 = \hat{s}_0) < 0 \} \nonumber \\
&=P\left \{\text{Re}\{U|h_0|\exp{\{j\phi_0\}}\Delta_{s,\hat{s}}^\ast\}< -\frac{\sqrt{G_0 E_0}}{2r_0^{\alpha_L}}|\Delta_{s,\hat{s}}|^2|h_0|^2 \right \} \nonumber \\
&=P\left \{\text{Re}\{U\}< -\frac{\sqrt{G_0 E_0}}{2r_0^{\alpha_L}}|\Delta_{s,\hat{s}}||h_0| \right \} \nonumber \\
&=F_{U_{\text{Re}}}\left (-\frac{\sqrt{G_0 E_0}}{2r_0^{\alpha_L}}|\Delta_{s,\hat{s}}||h_0|\right ) \label{eq:PEPcomp}
\end{align}
where $F_{U_{\text{Re}}}$ denotes the CDF of the RV $U_{\text{Re}}=\text{Re}\{U\}$.

Gil-Pelaez inversion theorem can be employed to compute the CDF $F_{U_{\text{Re}}}$ by using the CF of $U_{\text{Re}}$, $\Phi_{U_{\text{Re}}}(w)$, as follows \cite{Gil-Pelaez}:
\small
\begin{align}
&F_{U_{\text{Re}}}(u)= \frac{1}{2}-\frac{1}{\pi} \int_0^{\infty}\frac{\text{{\footnotesize Im}}\{\mathrm{e}^{-jwu}\Phi_{U_{\text{Re}}}(w)\}}{w}dw \nonumber \\
&\hspace{-0.7cm}=\!\frac{1}{2}\!-\!\frac{1}{\pi} \!\!\int_0^{\infty}\!\!\frac{\text{{\footnotesize Im}}\{({\footnotesize \cos(wu)}\!-\!j{\footnotesize \sin (wu)})(\text{{\footnotesize Re}}\{\Phi_{U_{\text{Re}}}(w)\})\!+\!j\text{{\footnotesize Im}}\{\Phi_{U_{\text{Re}}}(w)\})\}}{w}dw  \nonumber \\
&\hspace{-0.7cm}=\frac{1}{2}+\frac{1}{\pi} \int_0^{\infty}\frac{{\footnotesize \sin(wu)}\Phi_{U}(w)}{w}dw,   \label{eq:CDF}
\end{align}
\normalsize
where the last equation follows from the fact that the CF $\Phi_{U_{\text{Re}}}(w)$ is a real function, i.e., $\text{Im}\{\Phi_{U_{\text{Re}}}(w)\}=0$ and $\text{Re}\{\Phi_{U_{\text{Re}}}(w)\}=\Phi_{U_{\text{Re}}}(w)$, and $U(.)$ is a circularly symmetric RV, i.e., $\Phi_{U_{\text{Re}}}(w)=\Phi_{U}(w)$.

By inserting (\ref{eq:CDF}) into (\ref{eq:PEPcomp}), PEP can be written as
\small
\begin{equation}
\hspace{-0.2cm} {\footnotesize {P}}\{s_0 \! \! \to \! \! \hat{s}_0||h_0|,r_0 \}\! \! = \frac{1}{2}-\frac{1}{\pi} \int_0^{\infty}{\! \! \! \sin\left (\frac{\sqrt{G_0 E_0}}{2r_0^{\alpha_L}}|\Delta_{s,\hat{s}}||h_0|w\right )}w^{-1}\Phi_{U}(w)dw  \label{eq:PEP}.
\end{equation}
\normalsize
In (\ref{eq:PEP}), PEP is computed using the CF of the RV $U$. Since $U$ is the summation of two independent RVs, $I_{\agg}$ and $n$, $\Phi_{U}(w)$ is  equal to the product of the CFs of these two RVs, i.e., $\Phi_{U}(w)=\Phi_{I_{\agg}}(w) \Phi_{n}(w)$. $\Phi_{n}(w)=\exp\{-w^2(N_0/4)\}$ is the CF of a Gaussian RV \cite{Biglieri} and $\Phi_{I_{\agg}}(w)$ is calculated in the next subsection.

\subsection{Characteristic Function of the Aggregate Interference $I_{\agg}$}
Since $I_{\agg}$ is the sum of six independent PPPs as seen in (\ref{eq:6PPP}), using stochastic geometry, its CF can be written as
\begin{equation}
\Phi_{I_{\agg}}(w)=\prod_{G \in \{MM,Mm,mm\}}\Phi_{I_{\Psi_{\text{LOS}}}^{G}}(w)\Phi_{I_{\Psi_{\text{NLOS}}}^{G}}(w),  \label{eq:CFofIagg}
\end{equation}
where $\Phi_{I_{\Psi_{\text{LOS}}}^{G}}(w)$ and $\Phi_{I_{\Psi_{\text{LOS}}}^{G}}(w)$ are the CFs of LOS and NLOS components with antenna gain $G$.

Let $z_i=s_ih_i=|h_i|\exp{\{j(\theta_i+\phi_i)\}}$. The interference due to a LOS component with a generic antenna gain $G$ can be written as $I_{\Psi_{\text{LOS}}}^{G}=\sum_{i \in \Psi_{\text{LOS}}} \sqrt{G E_0}r_i^{-\alpha_L}z_i$. Then, its CF $\Phi_{I_{\Psi_{\text{LOS}}}^{G}}(w)=\mathbb{E}\{\exp \{jwI_{\Psi_{\text{LOS}}}^{G}\}\}$ can be obtained using the same steps as those in \cite{Haenggi} and can be expressed as in (\ref{eq:CFofILos1}) given on the next page.
\begin{figure*}
\begin{equation}
\Phi_{I_{\Psi_{\text{LOS}}}^{G}}(w)=\sum_{k=0}^{\infty}\frac{\exp\{-\lambda p_G \pi(R_B^2-r_0^2)\}[\lambda p_G\pi(R_B^2-r_0^2)]^k}{k!}\big(\mathbb{E}_{z_i,r_i}\{\exp\{jw\sqrt{G E_0}r_i^{-\alpha_L}z_i\}\})^k. \label{eq:CFofILos1}
\end{equation}
\vspace{-.5cm}
\end{figure*}
By using the Taylor series expansion for the exponential function, one can rewrite the equation in (\ref{eq:CFofILos1}) and further express it using similar steps as in \cite{Win} as follows:
\begin{align}
&\Phi_{I_{\Psi_{\text{LOS}}}^{G}}(w) \nonumber
\\
&=\exp\{\lambda p_G\pi(R_B^2-r_0^2)[-1+\mathbb{E}_{z_i,r_i}\{\exp\{jw\sqrt{G E_0}r_i^{-\alpha_L}z_i\}\}]\} \nonumber \\
&\stackrel{(a)}{=}\exp\left\{2\lambda p_G\pi\mathbb{E}_{z_i}\{\int_{r_0}^{R_B}(\exp\{jw\sqrt{G E_0}r_i^{-\alpha_L}z_i\}-1)\}r_idr_i\right\} \nonumber \\
&\stackrel{(b)}{=}\exp\left\{2\lambda p_G\pi\int_{r_0}^{R_B}(\Phi_{\mathbf{z}}(\sqrt{G E_0}wr_i^{-\alpha_L})-1)r_idr_i\right\} \nonumber \\
&\stackrel{(c)}{=}\exp\left\{2\lambda p_G\pi\int_{r_0}^{R_B}(\Phi_0(\sqrt{G E_0}|w|r_i^{-\alpha_L})-1)r_idr_i\right\} \nonumber \\
&\stackrel{(d)}{=}\exp\left\{-2\lambda p_G\pi\frac{(\sqrt{G E_0}|w|)^{2/\alpha_L}}{\alpha_L}\int_{\sqrt{G E_0}|w|R_B^{-\alpha_L}}^{\sqrt{G E_0}|w|r_0^{-\alpha_L}}\frac{1-\Phi_0(t)}{t^{2/\alpha_L+1}}dt\right\} \label{eq:CFofILos2}
\end{align}
where (a) follows from $r_i$ having a PDF of $2r_i/(R_B^2-r_0^2)$ if $r_0 \leq r_i \leq R_B$ and zero otherwise, (b) originates from the definition of the CF, (c) follows from the fact that $\mathbf{z}$ has a spherically symmetric (SS) PDF and its CF is also SS, i.e., $\Phi_{\mathbf{z}}(w)=\Phi_0(w)$ for some $\Phi_0(.)$, (d) is obtained by applying a change of variables with $t=\sqrt{G E_0}|w|r_i^{-\alpha_L}$.

$\Phi_0(t)$ can be found using the properties of an SS distribution:
\begin{align}
\Phi_0(t)=\Phi_{z_i}(t)&=\mathbb{E}\{\mathrm{e}^{jtz_i}\}\nonumber \\
&=\mathbb{E}_{x_i}\{\cos(tx_i)\}+j\underbrace{\mathbb{E}_{y_i}\{\sin(ty_i)\}}_0 \nonumber \\
&=\mathbb{E}_{x_i}\{\cos(tx_i)\}  \label{eq:Phi0}
\end{align}
where $x_i=\text{Re}\{z_i\}$, $y_i=\text{Im}\{z_i\}$ and the second term is zero because $\sin$ is an odd-symmetric function.

By inserting the result in (\ref{eq:Phi0}) into (\ref{eq:CFofILos2}) and taking the expectation operator $\mathbb{E}_{x_i}\{.\}$ outside, the integral inside the exponential function can be calculated as shown at the top of next page in (\ref{eq:intT}) using the result from \cite[Eq. (3.771.4)]{integrals}, where ${}_p F_q$ is the generalized hypergeometric function. Then, by inserting the result of the integral in (\ref{eq:intT}) into (\ref{eq:CFofILos2}) and applying similar steps as in \cite{marco}, we obtain the closed-form expression for $\Phi_{I_{\Psi_{\text{LOS}}}^{G}}(w)$ in (\ref{eq:CFofILos3}).
\begin{figure*}[!t]
\begin{align}
T_i&=\int_{0}^{\sqrt{G E_0}|w|r_0^{-\alpha_L}}\frac{1-\cos(tx_i)}{t^{2/\alpha_L+1}}-\int_{0}^{\sqrt{G E_0}|w|R_B^{-\alpha_L}}\frac{1-\cos(tx_i)}{t^{2/\alpha_L+1}} \nonumber \\
&=\frac{\alpha_{L}}{2}\left(\sqrt{G E_0}|w|\right)^{-2/\alpha{L}}\left[R_B^2-r_0^2+{}_1F_2\left(-\frac{1}{\alpha_L};\frac{1}{2},1-\frac{1}{\alpha_L};-\frac{G E_0|w|^2}{4r_0^{2\alpha_L}}x_i^2\right)-{}_1F_2\left(-\frac{1}{\alpha_L};\frac{1}{2},1-\frac{1}{\alpha_L};-\frac{G E_0|w|^2}{4R_B^{2\alpha_L}}x_i^2\right)\right]. \label{eq:intT}
\end{align}
\vspace{-.8cm}
\end{figure*}
\small
\begin{figure*}[!t]
\begin{align}
\Phi_{I_{\Psi_{\text{LOS}}}^{G}}(w)=&\exp\{\lambda p_G\pi (r_0^2-R_B^2)\} \nonumber
\\
&\times\exp\left\{-\lambda p_G\pi r_0^2{}_2 F_2\left(-\frac{1}{2},-\frac{1}{\alpha_L};\frac{1}{2},1-\frac{1}{\alpha_L};-\frac{G E_0|w|^2\sigma_0}{4r_0^{2\alpha_L}}\right)+\lambda\pG\pi R_B^2{}_2 F_2\left(-\frac{1}{2},-\frac{1}{\alpha_L};\frac{1}{2},1-\frac{1}{\alpha_L};-\frac{G E_0|w|^2\sigma_0}{4R_B^{2\alpha_L}}\right)\right\}. \label{eq:CFofILos3}
\end{align}
\vspace{-.5cm}
\end{figure*}
\normalsize

Similarly, a closed-form expression for the CF of the interference due to NLOS BSs, $\Phi_{I_{\Psi_{\text{LOS}}}^{G}}(w)$, can be obtained by changing the boundaries of the integral and replacing $\alpha_{L}$ with $\alpha_{N}$ in (\ref{eq:CFofILos2}). More specifically, since NLOS BSs lie outside of the ball, integral is evaluated from $R_B$ to infinity and the expression for $\Phi_{I_{\Psi_{\text{LOS}}}^{G}}(w)$ is found as shown in (\ref{eq:CFofINlos}) on the next page.
\begin{figure*}[!t]
\begin{align}
\Phi_{I_{\Psi_{\text{NLOS}}}^{G}}(w)&=\exp\{\lambda p_G\pi R_B^2\}\exp\left\{-\lambda p_G\pi R_B^2{}_2 F_2\left(-\frac{1}{2},-\frac{1}{\alpha_N};\frac{1}{2},1-\frac{1}{\alpha_N};-\frac{G E_0|w|^2\sigma_0}{4R_B^{2\alpha_N}}\right)\right\}.  \label{eq:CFofINlos}
\end{align}
\vspace{-.5cm}
\end{figure*}

Finally, a closed-form expression for the CF of the aggregate network interference, $\Phi_{I_{\agg}}(w)$, can be obtained by inserting equations (\ref{eq:CFofILos3}) and (\ref{eq:CFofINlos}) into (\ref{eq:CFofIagg}).

\subsection{Average Pairwise Error Probability}
In this section, APEP is computed by averaging PEP. Averaging can be performed by taking the integral of the conditional PEP over $|h_0|$ and $r_0$ as follows:
\small
\begin{align}
&P_{\avg}\{s_0 \to \hat{s}_0 \}=\mathbb{E}_{|h_0|,r_0}\{P\{s_0 \to \hat{s}_0||h_0|,r_0 \}\} \nonumber \\
&\hspace{-0.8cm}=\mathbb{E}_{|h_0|,r_0}\left\{\frac{1}{2}-\frac{1}{\pi} \int_0^{\infty}{\! \! \! \sin\left(\frac{\sqrt{G_0 E_0}}{2r_0^{\alpha_L}}|\Delta_{s,\hat{s}}||h_0|w\right)}w^{-1}\Phi_{U}(w)dw\right\} \nonumber \\
&\hspace{-0.8cm}\stackrel{(a)}{=}\frac{1}{2}-\frac{1}{\pi}\int_0^{\infty}\mathbb{E}_{r_0}\left\{\mathbb{E}_{|h_0|}\left\{\sin \left(\frac{\sqrt{G_0 E_0}}{2r_0^{\alpha_L}}|\Delta_{s,\hat{s}}||h_0|w \right)\right \}\Phi_{U}(w)\right\}w^{-1}dw,   \label{eq:APEP1}
\end{align}
\normalsize
where (a) follows from the fact that $\Phi_U(w)$ depends only on $r_0$ not $|h_0|$. Hence, the expectation over $|h_0|$ can be computed in closed-form by employing the PDF of Rayleigh distribution and calculating the resulting integral as \cite{integrals}
\begin{align}
&\mathbb{E}_{|h_0|}\left\{\sin\left(\frac{\sqrt{\G_0 E_0}}{2r_0^{\alpha_L}}|\Delta_{s,\hat{s}}||h_0|w\right)\right\} \nonumber \\
&=\int_0^{\infty}\sin\left(\frac{\sqrt{G_0 E_0}}{2r_0^{\alpha_L}}|\Delta_{s,\hat{s}}||h_0|w\right)\frac{2\nu}{\sigma_0}\exp\{-\frac{\nu^2}{\sigma_0}\}d\nu \nonumber \\
&=\sqrt{\pi}\frac{\sqrt{G_0 E_0\sigma_0}}{4r_0^{\alpha_L}}|\Delta_{s,\hat{s}}|w\exp\left\{-\frac{G_0 E_0\sigma_0}{16r_0^{2\alpha_L}}|\Delta_{s,\hat{s}}|^2w^2\right\}. \label{eq:Avgoveralpha}
\end{align}

By substituting (\ref{eq:Avgoveralpha}) and the PDF of $r_0$ (i.e., $f_{r_0}(\xi)$) into (\ref{eq:APEP1}), APEP can be expressed as follows:
\small
\begin{align}
\hspace{-1.4cm}P_{\avg}\{s_0 \! \to \! \hat{s}_0 \}\!=\!\frac{1}{2}\!-\!\frac{1}{\pi}&\int_0^{\infty}\!\!\!\int_0^{\infty}\!\!\!\sqrt{\pi}\frac{\sqrt{G_0 E_0\sigma_0}}{4\xi^{\alpha_L}}|\Delta_{s,\hat{s}}|w\exp\left\{-\frac{G_0 E_0\sigma_0}{16\xi^{2\alpha_L}}|\Delta_{s,\hat{s}}|^2w^2\right\} \nonumber \\
&2\pi \lambda \xi \exp\{-\pi \lambda \xi^2\}\Phi_U(w)w^{-1}d\xi dw. \label{eq:APEP2}
\end{align}
\normalsize

Finally, substituting $\Phi_U(w)$ into (\ref{eq:APEP2}) and after some algebraic manipulations, APEP can be rewritten as
\begin{figure*}[!t]
\begin{align}
P_{\avg}\{s_0\to\hat{s}_0\}= \frac{1}{2}-\sqrt{\pi}\lambda\frac{\sqrt{G_0\SNR}}{2}|\Delta_{s,\hat{s}}|\int_0^{\infty}\int_0^{\infty}
\exp\left\{-\frac{\SNR}{4\xi^{2\alpha_L}}|\Delta_{s,\hat{s}}|^2w^2\right\}\exp\{-\pi\lambda\xi^2\}\exp\{-w^2N_0/4\} \nonumber \\ \times \prod_{G \in \{MM,Mm,mm\}}[\exp\{\lambda p_G\pi\xi^2\}\exp\{-\lambda p_G\pi\xi^2 {}_2F_2(\xi,\alpha_L)+\lambda p_G\pi R_B^2[{}_2F_2(R_B,\alpha_L)-{}_2F_2(R_B,\alpha_N)]\}]^{p_{G}} d\xi dw. \label{eq:APEP3}
\end{align}
\vspace{-.3cm}
\end{figure*}
shown at the top of next page in (\ref{eq:APEP3}), where we define $\SNR=E_0\sigma_0/4$ and
\begin{equation}
{}_2F_2(x,y)={}_2F_2\left(\frac{1}{2},-\frac{1}{y};\frac{1}{2},1-\frac{1}{y};-\frac{G \SNR|w|^2}{x^{2y}}\right).
\end{equation}

\subsection{Average Symbol Error Probability}
In this section, we approximate ASEP from APEP in (\ref{eq:APEP3}) by using NN approximation. The advantage of the NN approximation is that it only depends on the minimum distance in the constellation and the number of nearest neighbors \cite{Goldsmith}. In Section II, $|\Delta_{s,\hat{s}}|$ is defined as the distance between the constellation points $s_0$ and $\hat{s}_0$. Hence, we define $\Delta_{min}$ as the distance of $s_0$ to its nearest neighbors, i.e., $\Delta_{min}=\min_{\hat{s}_0 \ne s_0}|\Delta_{s_0,\hat{s}_0}|$. Also, let $k_{d_{\min}}$ denote the number of nearest neighbors of $s_0$ having distance $\Delta_{min}$. Now, ASEP can be approximated as
\begin{equation}
\text{ASEP} \approx  k_{d_{\min}} P_{\avg}\{\Delta_{min}\}.
\end{equation}
where $k_{d_{\min}}=2$ when modulation order ($\gamma$) is greater than 2, and $\Delta_{min}=2\sin(\pi/\gamma)$ assuming multilevel phase shift keying (MPSK) modulation.

\section{ASEP in the Presence of Beamsteering Errors}
In Section \ref{sec:error_prob_analysis}, MU and the serving BS are assumed to be aligned perfectly and ASEP is calculated in the absence of beamsteering errors. However, in practice, it may not be easy to have perfect alignment. Therefore, in this section, we investigate the effect of beamforming alignment errors on ASEP. We employ an error model similar to that in \cite{Wildman}. Let $|\epsilon|$ be the random absolute beamsteering error of the MU's beam toward the serving BS with zero-mean and bounded absolute error $\epsilon_{\text{max}} \le \pi$. It is appropriate to consider the absolute beamsteering error due to symmetry in the gain $\G$. The PDF of the effective antenna gain $\G$ with alignment error can be explicitly written as \cite{Marco2}
\begin{align}
\hspace{-.8cm}f_{G}(\g)&=F_{|\epsilon|}\left(\frac{\theta}{2}\right)^2\delta(\g-MM)\nonumber \\
&+2F_{|\epsilon|}\left(\frac{\theta}{2}\right)\!\!\left(\!\!1-F_{|\epsilon|}\left(\frac{\theta}{2}\right)\!\!\right)\delta(\g-Mm) \nonumber \\
&+\left(1-F_{|\epsilon|}\left(\frac{\theta}{2}\right)\right)^2\delta(\g-mm),
\label{eq:PDFofG}
\end{align}
where $\delta(\cdot)$ is the Kronecker's delta function, $F_{|\epsilon|}(x)$ is the CDF of the misalignment error and (\ref{eq:PDFofG}) follows from the definition of CDF, i.e., $F_{|\epsilon|}(x)=P\{|\epsilon|\le x\}$. Assume that the error is distributed according to a Gaussian distribution, so absolute error $|\epsilon|$ follows a half normal distribution and $F_{|\epsilon|}(x)=\text{erf}(x/(\sqrt{2}\sigma_{\BE}))$, where $\text{erf}(\cdot)$ denotes the error function.

From (\ref{eq:APEP3}), it is clear that PEP depends on the effective antenna gain between the MU and the serving BS, and so does the ASEP. Thus, PEP can be calculated by averaging over the distribution of $G$, $f_{G}(\g)$, as follows:
\begin{align}
P_{\avg}\{s_0\to\hat{s}_0\}&= \mathbb{E}_{G}\{P_{\avg}\{s_0\to\hat{s}_0;G\}\} \nonumber \\
&=\int_0^{\infty}P_{\avg}\{s_0\to\hat{s}_0;\g\}f_{G}(\g)d \g \nonumber \\
&=(F_{|\epsilon|}(\theta/2))^2 P_{\avg}\{s_0\to\hat{s}_0;MM\} \nonumber \\
&+2(F_{|\epsilon|}(\theta/2))\bar{F}_{|\epsilon|}(\theta/2) P_{\avg}\{s_0\to\hat{s}_0;Mm\} \nonumber \\
&+\bar{F}_{|\epsilon|}(\theta/2)^2 P_{\avg}\{s_0\to\hat{s}_0;mm\},
\end{align}
where we define $\bar{F}_{|\epsilon|}(\theta/2)=1-F_{|\epsilon|}(\theta/2)$.

\section{Numerical Results}\label{sec:num_results}
In this section, we provide numerical results to evaluate the error performance of a downlink mmWave cellular network. In all figures, LOS and NLOS path loss exponents are $\alpha_L=2.1$ and $\alpha_N=4$, respectively. In the non-mmW case, all BSs are assumed to be LOS and the path loss component is equal to 2.1. Also, the radius of the LOS ball $R_B$ is assumed to be equal to 141 meters similarly as in \cite{Bai2}.

First, we compare the performance of the mmWave network with that of the non-mmWave network (antennas are omnidirectional, and all BSs are LOS). In Fig. 1, ASEP versus SNR is plotted for different BS densities with BPSK modulation. As shown in Fig. 1, we have better error performance in the mmWave scenario than in the non-mmWave one. Also, with the increasing BS density, ASEP is decreasing.
\begin{figure}
  \centering
  \includegraphics[width=\figsize\textwidth]{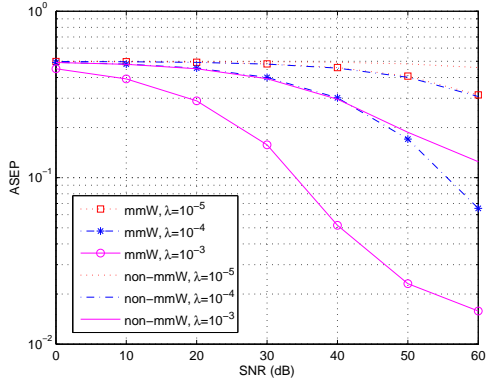}
\caption{\footnotesize ASEP as a function of the $\text{SNR}=E_0\sigma_0/4$ for different BS densities $\lambda$ ($\alpha_L=2.1$, $\alpha_N=4$, $M=10$ dB, $m=-10$ dB, $\theta=15$, BPSK \normalsize)}
\vspace{-.5cm}
\end{figure}

Next, we plot the ASEP with different antenna main lobe gains and different BS densities. The numerical results in Fig. 2 show that with increasing main lobe gain $M$, ASEP decreases significantly. Also, note that different combinations of main lobe gain and BS density, e.g. ($M=20dB$, $\lambda=10^{-5}$) and ($M=10dB$, $\lambda=10^{-4}$) lead to the same error performance. Hence, the same error performance can be achieved by either decreasing BS density while increasing the main lobe gain, or vice versa.
\begin{figure}
  \centering
  \includegraphics[width=\figsize\textwidth]{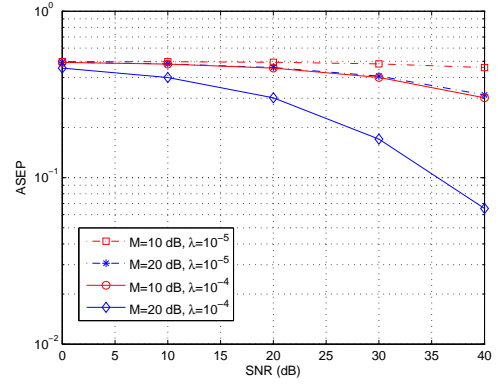}
\caption{\footnotesize ASEP as a function of the $\text{SNR}=E_0\sigma_0/4$ for different antenna main lobe gains $M$ and BS densities $\lambda$ ($\alpha_L=2.1$, $\alpha_N=4$, $m=-10$ dB, $\theta=15$, BPSK \normalsize)}
\end{figure}

In Fig. 3, we also compare ASEP for different modulation orders assuming MPSK modulation. As the modulation order increases, the minimum distance between the nearest neighbors decreases. Thus, as expected, error performance of the network gets worse with the increase in modulation size.
\begin{figure}
  \centering
  \includegraphics[width=\figsize\textwidth]{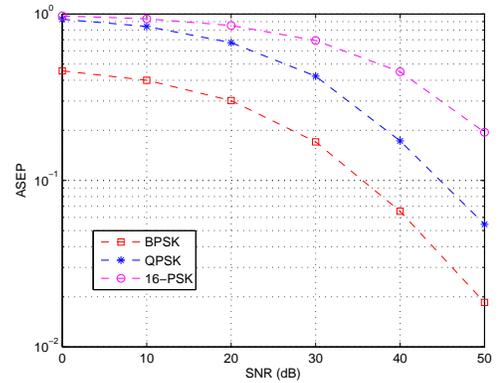}
\caption{\footnotesize ASEP as a function of the $\text{SNR}=E_0\sigma_0/4$ for different modulation orders Mo ($\alpha_L=2.1$, $\alpha_N=4$, $M=20$, dB $m=-10$ dB, $\theta=15$, $\lambda=10^{-4}$\normalsize)}
\vspace{-.6cm}
\end{figure}

Finally, the effect of beamsteering errors on the error performance is analyzed in Fig. 4. ASEP versus SNR is plotted for different standard deviations of the alignment error. As can be seen, ASEP is getting worse with the degradation in the alignment angle. $\sigma_{\BE}=0$ corresponds to the case with no alignment error and it has the best error performance as expected. Since $\sigma_{\BE}=2$ has the same ASEP with $\sigma_{\BE}=0$, we can infer that the alignment error until $\sigma_{\BE}=2$ can be tolerated and ASEP increases significantly for $\sigma_{\BE} > 2$.
\begin{figure}
  \centering
  \includegraphics[width=\figsize\textwidth]{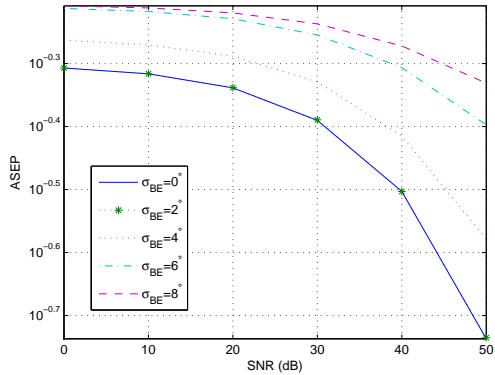}
\caption{\footnotesize ASEP as a function of the $\text{SNR}=E_0\sigma_0/4$ in the presence of beamsteering error for different standard deviations of alignment error ($\alpha_L=2.1$, $\alpha_N=4$, $M=20$, dB $m=-10$ dB, $\theta=15$, $\lambda=10^{-5}$, BPSK \normalsize)}
\vspace{-.8cm}
\end{figure}

\section{Conclusion}\label{sec:conclusion}
In this paper, we have analyzed the average error performance of downlink mmWave cellular networks, incorporating the distinguishing features of mmWave communication into the average error probability analysis. Sectored antenna and simplified ball-LOS models have been considered to simplify the analysis. Numerical results show that employing directional antennas improves the error performance. Also, we show that better ASEP values can be obtained by increasing BS density and main lobe gain. Investigating the effects of using different LOS probability functions instead of the simplified ball-LOS model, and incorporating more general fading models remain as future work.



\begin{thebibliography}{99}
\bibitem{UMTS} ``Mobile traffic forecasts: 2010-2020 report," in \emph{Proc. UMTS Forum Rep.}, vol. 44, pp. 1-92, Zurich, Switzerland, 2011.

\bibitem{Ericsson} Ericson, ``More than 50 billion connected devices," Feb. 2011.

\bibitem{Rappaport} T. Rappaport \emph{et al.}, ``Millimeter wave mobile cimmunications for 5G cellular: It will work!," \emph{IEEE Access}, vol. 1, pp. 335-349, May 2013.

\bibitem{Andrews} J. G. Andrews \emph{et al.}, ``What will 5G be?," \emph{IEEE J. Sel. Areas Commun.}, vol. 32, no. 6, pp. 1065-1082, Jun. 2014.

\bibitem{Rangan} S. Rangan and T.S. Rappaport and E. Erkip, ``Millimeter-wave cellular wireless networks: Potentials and challenges," \emph{Proceedings of the IEEE}, vol. 102, no. 3, pp. 366-385, Mar. 2014.

\bibitem{Ghosh} A. Ghosh and T. A. Thomas and M. C. Cudak and R. Ratasuk and P. Moorut and F. W. Vook and T. S. Rappaport and G. R. MacCartney and Shu Sun and Shuai Nie, ``Millimeter Wave Enhanced Local Area Systems: A
high data rate approach for future wireless networks," \emph{IEEE Journal on Sel. Areas in Comm.}, Special Issue on 5G, Jul., 2014.

\bibitem{Bai2} T. Bai and R. W. Heath, ``Coverage and rate analysis for millimeter wave
cellular networks," \emph{IEEE Trans. Wireless Commun.}, 2014. Submitted, available at http://arxiv.org/abs/1402.6430.

\bibitem {Marco2} M. Di Renzo ``Stochastic Geometry Modeling and Analysis of Multi-Tier Millimeter Wave Cellular Networks," 2014. Submitted, available at http://arxiv.org/abs/1410.3577.

\bibitem{Marco4} M. Di Renzo and P. Guan, ``Stochastic geometry modeling of coverage and rate of cellular networks using the Gil-Pelaez inversion theorem", \emph{IEEE Commun. Lett.}, vol. 18, no. 9, pp. 1575-1578, Sep. 2014.

\bibitem{Baccelli} F. Baccelli and B. Blaszczyszyn, ``Stochastic Geometry and Wireless
Networks, Part I: Theory, Part II: Applications,"  NOW: Foundations and Trends in Networking, 2010.

\bibitem{Win} M. Z. Win and P. C. Pinto and L. A. Shepp, ``A mathematical theoory of network interference and its applications," \emph{Proc. of the IEEE}, vol. 97, no. 2, pp. 205-230, Feb. 2009.

\bibitem{marco} P. Guan and M. Di Renzo, ``Stochastic geometry analysis of the average error probability of downlink cellular networks," \emph{IEEE Int. Conf. Comput. Netw. Commun.}, pp. 649-655, Feb. 2014.

\bibitem{Simon} M. K. Simon and M. S. Alouini, \emph{Digital Communication over Fading Channels}, 2nd ed. Hoboken, NJ, USA: Willey, 2005.

\bibitem{Bai3} A. Thornburg and T. Bai and R. W. Heath, ``Performance analysis of mmWave ad hoc networks," \emph{IEEE Trans. Signal Process.}, 2014. Submitted, available at http://arxiv.org/abs/1412.0765.

\bibitem{Andrews2} Jeffrey G. Andrews and François Baccelli and Radha Krishna Ganti, ``A Tractable approach to
coverage and rate in cellular networks," \emph{IEEE Trans. Commun.}, vol. 59, no. 11, Nov. 2011.

\bibitem{Hunter} A. Hunter and J. Andrews and S. Weber, ``Transmission capacity of ad hoc networks with spatial diversity," \emph{IEEE Trans. Wireless Commun.}, vol. 7, no. 12, pp. 5058-5071, Dec. 2008.

\bibitem{Bai1} T. Bai and R. Vaze and R. W. Heath Jr., ``Analysis of blockage effects on urban cellular networks," \emph{IEEE Trans. Wireless Commun.}, vol. 13, no. 9, pp. 5070-5083, Sept. 2014.

\bibitem{Samoradnitsky} G. Samoradnitsky and M.S. Taqqu, \emph{Stable Non-Gaussian Random Processes}, Chapman and Hall, Jun. 1994.

\bibitem{Gil-Pelaez} J. Gil-Pelaez, ``Notes on the inversion theorem," \emph{Biometrika}, vol. 38, pp. 481-482, Dec. 1951.

\bibitem{Biglieri} E. Biglieri and G. Caire and G. Taricco and J. Ventura-Traveset, ``Computing error probabilities over fading channels: A unified approach," \emph{European Trans. Telecommun.}, vol. 9, no. 1, pp. 15-25, Jan.-Feb. 1998

\bibitem{Haenggi} M. Haenggi and R. K. Ganti, ``Interference in Large Wireless Networks," Foundations and Trends in Networking, vol. 3, no. 2, pp. 127-248, 2009.

\bibitem{integrals} I. S. Gradshteyn and I. M. Ryzhik, ``Tables of Integrals, Series, and Products," San Diego, CA: Academic, 7th ed., 2007.

\bibitem{Goldsmith} A. Goldsmith, \emph{Wireless Communications}, Cambridge University Press, 2005.

\bibitem{Wildman} J. Wildman, P. H. J. Nardelli, M. Latva-aho, S. Weber, ``On the Joint Impact of Beamwidth and Orientation Error on Throughput in Directional Wireless Poisson Networks," \emph{to appear in the IEEE Trans. Wireless Commun.}, 2014.


\end{thebibliography}
\end{document}